\documentclass[review,12pt]{elsarticle}

\usepackage{amssymb}
\usepackage{amsthm}
\usepackage{amsmath}
\usepackage{mathrsfs}
\usepackage{graphicx}
\usepackage{epstopdf}
\usepackage{float}
\usepackage{caption}
\usepackage{soul}
\usepackage{color,soul} 
\usepackage{color} 
\usepackage{subcaption}
\usepackage{bm}
\usepackage{bbm}
\usepackage{mathrsfs}
\usepackage{cleveref}
\usepackage{soul}
\biboptions{sort&compress}

\journal{Computer Methods in Applied Mechanics and Engineering}

\makeatletter
\def\@author#1{\g@addto@macro\elsauthors{\normalsize%
    \def\baselinestretch{1}%
    \upshape\authorsep#1\unskip\textsuperscript{%
      \ifx\@fnmark\@empty\else\unskip\sep\@fnmark\let\sep=,\fi
      \ifx\@corref\@empty\else\unskip\sep\@corref\let\sep=,\fi
      }%
    \def\authorsep{\unskip,\space}%
    \global\let\@fnmark\@empty
    \global\let\@corref\@empty  
    \global\let\sep\@empty}%
    \@eadauthor={#1}
}
\makeatother

\makeatletter
\def\thickhline{%
  \noalign{\ifnum0=`}\fi\hrule \@height \thickarrayrulewidth \futurelet
   \reserved@a\@xthickhline}
\def\@xthickhline{\ifx\reserved@a\thickhline
               \vskip\doublerulesep
               \vskip-\thickarrayrulewidth
             \fi
      \ifnum0=`{\fi}}
\makeatother

\newlength{\thickarrayrulewidth}
\begin{document}

\begin{frontmatter}



\title{Phase field modelling of fracture and fatigue in Shape Memory Alloys}


\author{Marlini Simoes\fnref{Cam}}

\author{Emilio Mart\'{\i}nez-Pa\~neda\corref{cor1}\fnref{IC}}
\ead{e.martinez-paneda@imperial.ac.uk}

\address[Cam]{Cavendish Laboratory, University of Cambridge, Cambridge CB3 0HE, UK}

\address[IC]{Department of Civil and Environmental Engineering, Imperial College London, London SW7 2AZ, UK}

\cortext[cor1]{Corresponding author.}

\begin{abstract}
We present a new phase field framework for modelling fracture and fatigue in Shape Memory Alloys (SMAs). The constitutive model captures the superelastic behaviour of SMAs and damage is driven by the elastic and transformation strain energy densities. We consider both the assumption of a constant fracture energy and the case of a fracture energy dependent on the martensitic volume fraction. The framework is implemented in an implicit time integration scheme, with both monolithic and staggered solution strategies. The potential of this formulation is showcased by modelling a number of paradigmatic problems. First, a boundary layer model is used to examine crack tip fields and compute crack growth resistance curves (R-curves). We show that the model is able to capture the main fracture features associated with SMAs, including the toughening effect associated with stress-induced phase transformation. Insight is gained into the role of temperature, material strength, crack density function and fracture energy homogenisation. Secondly, several 2D and 3D boundary value problems are addressed, demonstrating the capabilities of the model in capturing complex cracking phenomena in SMAs, such as unstable crack growth, mixed-mode fracture or the interaction between several cracks. Finally, the model is extended to fatigue and used to capture crack nucleation and propagation in biomedical stents, a paradigmatic application of nitinol SMAs.\\

\end{abstract}


\end{frontmatter}


\section{Introduction}
\label{Introduction}

\begin{figure}[H]
  \makebox[\textwidth][c]{\includegraphics[width=1.1\textwidth]{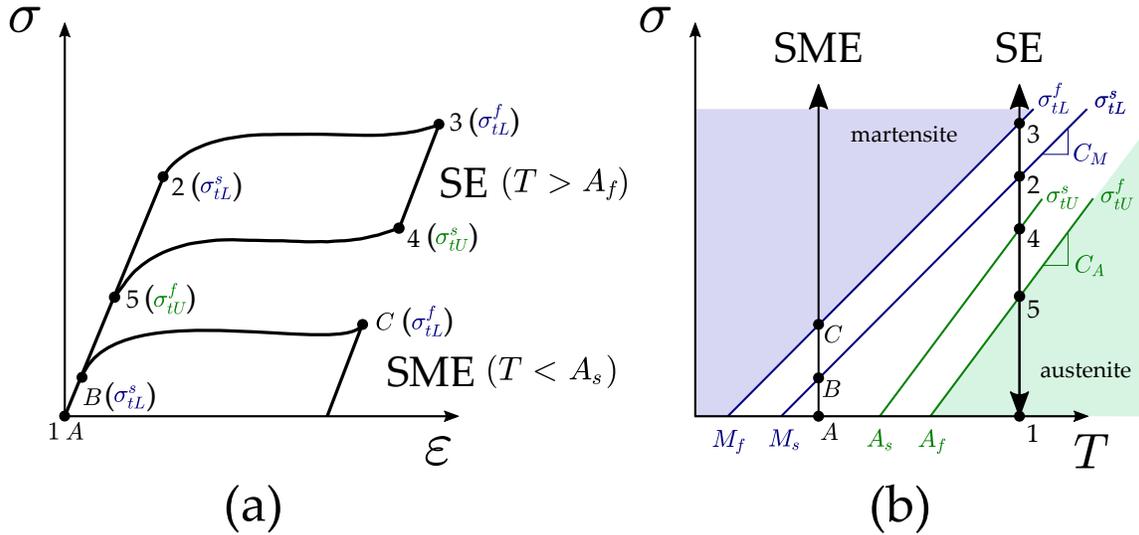}}%
  \caption{Superelasticity (SE) and shape memory effect (SME) phenomena in SMAs: (a) representative stress versus strain curves, and (b) representative stress versus temperature curves.}
  \label{fig:SMASketch}
\end{figure}

Shape Memory Alloys (SMAs) have gained increasing attention in recent years, with applications spanning the areas of aerospace, bioengineering, transport and infrastructure, among others \cite{Lagoudas2008}. The popularity of these smart, multi-functional materials is largely grounded on their capacity to sustain notably large recoverable strains (up to 10\%) as a result of transformation between their austenitic and martensitic phases. This transformation can be attained by changing the mechanical load (stress-induced transformation), the temperature (temperature-induced transformation) or both (stress and temperature-induced transformation). As shown in Fig. \ref{fig:SMASketch}a, two phenomena are intrinsic to SMAs and their phase transformation: superelasticity (SE) and shape memory effect (SME). In both cases, due to the creation of a stress-induced phase, the application of a mechanical load renders a non-linear response with very large strains. However, only superelastic alloys, such as nickel-titanium (nitinol), can achieve full strain recovery when the load is removed, exhibiting a hysteresis loop in the stress versus strain response. SMAs experiencing the shape memory effect show a large residual strain after unloading and require a change of temperature to recover their original shape. 

In SMAs where both superelasticity and shape memory effect can take place, their occurrence is governed by the temperature of the system, $T$. Fig. \ref{fig:SMASketch}b shows a typical SMA stress-temperature curve, where $M_f$, $M_s$, $A_s$ and $A_f$ denote the martensite end and start temperatures, and the austenite start and end temperatures, respectively. If $T>A_f$ (point 1), superelasticity will be observed; upon loading, the material will be in the austenite phase until reaching the stress level dictating the start of the transformation $\sigma_{tL}^s$ (point 2), and the transformation will end upon attaining $\sigma_{tL}^f$ (point 3), when the material will be fully martensitic. If the load is removed, full recovery is observed; reverse transformation starts at $\sigma_{tU}^s$ (point 4) and all the martensite transforms into austenite upon reaching $\sigma_{tU}^f$ (point 5). The shape memory effect will be observed if $T<A_s$; austenite to martensite transformation will start when the stress level reaches the threshold $\sigma_{tL}^s$ (point B), ending upon attaining $\sigma_{tL}^f$ (point C). However, no reverse transformation takes place upon unloading since austenite is not stable at this temperature; $T$ must be raised to eliminate the residual strains. The magnitude of the temperature-dependent stress thresholds ($\sigma_{tL}^s$, $\sigma_{tL}^f$, $\sigma_{tU}^s$, $\sigma_{tL}^f$) is governed by $C_M$ and $C_A$, the slope of the stress-temperature diagram for martensite and austenite, respectively. The material behaviour in the phase transformation region is typically defined as a function of the martensitic volume fraction, $\xi$ ($0 \leq \xi \leq 1$). The reader is referred to Refs. \cite{Auricchio1997,Taylorb1997,Patoor2006,Freed2007} for more details.\\

The fracture and fatigue behaviour of SMAs has attracted significant interest, from both numerical and experimental perspectives; see, for example, the reviews by Robertson \textit{et al.} \cite{Robertson2012} and Baxevanis and Lagoudas \cite{Baxevanis2015}. Since the yield stress of SMAs is typically much larger than the transformation stress $\sigma_{tL}^f$ \cite{McKelvey2001}, a stress-induced transformation zone develops in the vicinity of the crack tip. This crack tip transformation region has been characterised using infrared (IR) thermography \cite{Gollerthan2009} and synchrotron X-ray diffraction \cite{Robertson2007a,Daymond2007}. The high stresses of the transformation region dominate the crack growth resistance of SMAs and result in energy dissipation and material toughening \cite{Robertson2007,Haghgouyan2019}. Finite element models have been developed to predict the role of transformation toughening and reverse transformation on crack propagation \cite{Freed2007,Baxevanis2013,Baxevanis2014,Karimi2019}. While important insight has been gained, these efforts have been restricted to discrete numerical methods, such as cohesive zone formulations. Discrete numerical methods for fracture are limited when dealing with the complex conditions of practical applications and, consequently, important challenges remain unaddressed (crack nucleation, mixed-mode, interacting cracks, etc.). Moreover, to the best of the authors' knowledge, a modelling framework capable of explicitly predicting fatigue crack growth behaviour in SMAs has not been presented yet. The main objective of this study is to address these two important knowledge gaps by developing a phase field-based computational framework for fracture and fatigue cracking in SMAs.\\

The phase field method has proven to be a compelling variational framework for predicting advanced fracture problems. The classical Griffith fracture energy balance \cite{Griffith1920} is revisited as an energy minimisation problem by solving for an auxiliary variable, the phase field parameter $\phi$ \cite{Francfort1998,Bourdin2000}. This enables capturing, on the original finite element mesh, complex cracking phenomena such as crack nucleation, branching, kinking or merging in arbitrary geometries and dimensions (see, e.g., \cite{Borden2012,Borden2016,McAuliffe2016,TAFM2020}). Not surprisingly, the method is enjoying great popularity and its success has been extended to numerous applications. Recent examples include fracture of functionally graded materials \cite{CPB2019,DT2020}, composites delamination \cite{Quintanas-Corominas2019,Quintanas-Corominas2020a}, cracking in solar-grade silicon \cite{Paggi2018}, hydrogen embrittlement \cite{CMAME2018,Wu2020b,CS2020}, rock fracture \cite{Zhou2019b,Schuler2020}, and fatigue damage \cite{Lo2019,Carrara2020}, among others; see \cite{Wu2020} for a review.\\

In this work, we present the first phase field formulation for fracture (and fatigue) in Shape Memory Alloys (SMAs). The constitutive behaviour of the solid includes both stress and temperature-induced phase transformations, capturing the superelasticity (reverse transformation) and shape memory effects. The evolution of the phase field variable is driven by the combination of elastic and transformation strain energy densities. Two different phase field dissipation functions are considered, corresponding to the so-called AT1 \cite{Pham2011} and AT2 \cite{Ambrosio1991,Bourdin2000} models, and their influence is assessed. The model is implemented in an implicit time integration scheme and the coupled displacement-phase field problem is solved using both monolithic (quasi-Newton) and staggered schemes, demonstrating the robustness of the framework. Several paradigmatic boundary value problems are addressed to demonstrate the potential of the framework in providing physical insight into the fracture behaviour of SMAs and modelling complex, large scale 3D fatigue problems. The remainder of the paper is organized as follows. The theoretical framework is presented in Section \ref{Sec:Theory}. Then, the finite element implementation is described in Section \ref{Sec:NumModel}. Representative numerical results are shown in Section \ref{Sec:FEMresults}. First, a boundary layer model is used to gain insight into stationary and propagating cracks. Secondly, we proceed to model mode I fracture in a square plate, a paradigmatic benchmark in phase field fracture. Mixed-mode conditions and crack coalescence is then investigated using an asymmetric double-notched bar. Finally, the results section concludes with a 3D large scale analysis of fatigue failure of a biomedical stent. Concluding remarks end the paper in Section \ref{Sec:ConcludingRemarks}.

\section{A phase field fracture formulation for Shape Memory Alloys}
\label{Sec:Theory}

\subsection{Constitutive behaviour of SMAs}
\label{sec:ConstitutiveSMA}

To constitutively describe the material behaviour of SMAs we follow the so-called unified model by Lagoudas and co-workers \cite{Boyd1996,Lagoudas1996,Lagoudas2008}, including recent extensions to capture gradual phase transformations, and the stress-dependencies of the inelastic recoverable strain and the phase diagram slope \cite{Hartl2010,Lagoudas2012}. The model builds upon the rule of mixtures to determine the magnitude of relevant properties for material points located in the phase transformation region. Thus, the effective value of any relevant phase-dependent parameter ($\Theta$) is a function of the martensitic volume fraction $\xi$ ($0 \leq \xi \leq 1$) and its magnitude in the austenitic ($\Theta^A$) and martensitic ($\Theta^M$) phases;
\begin{equation}\label{eq:rulemixtures}
     \Theta \left( \xi \right) = \left( 1 - \xi \right) \Theta^A  + \xi \Theta^M.
\end{equation}

\subsubsection{Thermodynamic potential}

A Gibbs free energy potential $\mathcal{G}$ can be defined as a function of the terms corresponding to the austenitic ($\mathcal{G}^A$) and martensitic ($\mathcal{G}^M$) phases, the martensitic volume fraction ($\xi$), and a mixing term due to the phase transformation ($\mathcal{G}^{mix}$). Both $\mathcal{G}^A$ and $\mathcal{G}^M$ are functions of the Cauchy stress tensor $\bm{\sigma}$ and the absolute temperature $T$, while $\mathcal{G}^{mix}$ is a function of $\bm{\sigma}$, the transformation strain tensor $\bm{\varepsilon}^t$, and the so-called transformation hardening energy $g^t$, which is a measure of the nonlinear change in the mixing energy as transformation progresses at constant stress \cite{Lagoudas2012}. Thus, the Gibbs energy reads,
\begin{equation}
    \mathcal{G} \left( \bm{\sigma}, \, T, \, \bm{\varepsilon}^t, \, \xi , \, g^t \right) = \left( 1 - \xi \right) \mathcal{G}^A \left( \bm{\sigma}, T \right) + \xi \mathcal{G}^M \left( \bm{\sigma}, \, T \right) + \mathcal{G}^{mix} \left( \bm{\sigma}, \, \bm{\varepsilon}^t, \, g^t \right).
\end{equation}

Assuming a quadratic dependence on stress \cite{Lagoudas2012}, for a material with density $\rho$, thermal expansion tensor $\bm{\alpha}$, compliance tensor $\mathcal{\bm{S}}$, and specific heat $c$, the $\mathcal{G}^A$ ($\gamma=A$) and $\mathcal{G}^M$ ($\gamma=M$) terms read,
\begin{align}
    \mathcal{G}^\gamma \left( \bm{\sigma}, T \right)  = - \frac{1}{2 \rho} & \bm{\sigma} : \bm{S}^\gamma : \bm{\sigma}  - \frac{1}{\rho} \bm{\sigma} : \bm{\alpha} \left( T - T_0 \right)  \\ \nonumber
    & + c^\gamma \left[ \left( T- T_0 \right) - T \ln \left( \frac{T}{T_0} \right) \right] - s_0^\gamma T + u_0^\gamma ,
\end{align}

\noindent where $s_0$, $u_0$, and $T_0$ respectively denote the specific entropy, specific internal energy and temperature at the reference state. Finally, the mixing term of the Gibbs free energy is given by,
\begin{equation}
    \mathcal{G}^{mix} \left( \bm{\sigma}, \bm{\varepsilon}^t, g^t \right) = - \frac{1}{\rho} \bm{\sigma} : \bm{\varepsilon}^t + \frac{g^t}{\rho} .
\end{equation}

\subsubsection{Evolution equations}

Evolution equations for the transformation strain $\bm{\varepsilon}^t$ and the hardening variable $g^t$ are now provided. $\bm{\varepsilon}^t$, an inelastic strain tensor generated during transformation from austenite to martensite, is a function of the rate of the martensitic volume fraction $\dot{\xi}$ and the so-called transformation direction tensor $\bm{\Lambda}^t$:
\begin{equation}\label{eq:flowrule}
    \dot{\bm{\varepsilon}}^t = \bm{\Lambda}^t \dot{\xi}.
\end{equation}

\noindent If the rate of the martensitic volume fraction is positive ($\dot{\xi}>0$), forward transformation takes place ($\bm{\Lambda}^t = \bm{\Lambda}^t_{fwd}$) and the transformation direction tensor equals 
\begin{equation}
    \bm{\Lambda}^t_{fwd} = \frac{3}{2} H^{cur} \frac{\bm{\sigma}'}{\sigma_e},
\end{equation}

\noindent where the prime symbol $'$ denotes deviatoric quantities, $\sigma_e$ is an effective stress, defined as in von Mises plasticity $\sigma_e=\sqrt{(3/2) \bm{\sigma}':\bm{\sigma}'}$, and $H^{cur}$ is the uniaxial transformation strain magnitude for complete transformation. On the other hand, if $\dot{\xi}<0$, then the transformation direction tensor corresponds to its reverse form ($\bm{\Lambda}^t = \bm{\Lambda}^t_{rev}$), defined as a function of the transformation strain and the martensite volume fraction at the reversal:
\begin{equation}
   \bm{\Lambda}^t_{rev} = \frac{\bm{\varepsilon}^{t-r}}{\xi^r}.
\end{equation}

The modelling framework has the capability of capturing the stress sensitivity of the maximum transformation strain. This is achieved by defining $H^{cur}$ as a decaying exponential function when the effective stress exceeds a critical quantity $\sigma_c$ \cite{Hartl2010}. Thus, $H^{cur}$ is given by:
\begin{equation}\label{eq:Hmaxmin}
    H^{cur} \left( \bar{\sigma} \right) =   \begin{cases}
    H_{min}      & \quad \text{if } \sigma_e \leq \sigma_c \\
    H_{min} + \left\{ H_{max} - H_{min} \left[ 1 - \exp \left( k\sigma_c -  k \sigma_e \right) \right]  \right\} & \quad \text{if } \sigma_e > \sigma_c
  \end{cases},
\end{equation}

\noindent where $H_{min}$ corresponds to the observable uniaxial two-way shape memory effect (TWSME), $k$ is a fitting parameter and $H_{max}$ is the ultimate transformation strain under uniaxial loading.\\

It remains to define an evolution equation for the transformation hardening energy variable $g^t$; this is achieved by means of a hardening function $f^t$, which takes distinct values during forward and reverse transformation. Thus,
\begin{equation}
    \dot{g}^t = f^t \dot{\xi},
\end{equation}

\noindent with the hardening function being of the form,
\begin{equation}
    f_{fwd}^t \left( \xi \right)= \frac{a_1}{2}  \left( 1 + \xi^{n_1} - \left( 1 - \xi \right)^{n_2} \right) + a_3
\end{equation}

\noindent when $\dot{\xi}>0$ ($f^t=f^t_{fwd}$); while for reverse transformation ($\dot{\xi}<0$, $f^t=f^t_{rev}$) the hardening function reads:
\begin{equation}
    f_{rev}^t \left( \xi \right) = \frac{a_2}{2}  \left( 1 + \xi^{n_3} - \left( 1 - \xi \right)^{n_4} \right) - a_3.
\end{equation}

\noindent Here, the exponents $n_1$, $n_2$, $n_3$ and $n_4$ take real numbers in the range $(0, 1]$ and are aimed at enabling the simulation of gradual hardening behaviour during transformation. The richer description provided, relative to other existing models (exponential, trigonometric, linear, etc.), enables matching the experimental data more closely \cite{Hartl2010}. Apart from that, the constants $a_1-a_3$ are computed from the common SMA material parameters $M_s$, $M_f$, $A_s$, $A_f$, $C_A$, and $C_M$, as described in Ref. \cite{Hartl2010}.

\subsubsection{Thermodynamically-consistent constitutive prescriptions}

A suitable free energy can be defined building upon a thermodynamically-consistent energy imbalance, as elaborated elsewhere \cite{Lagoudas2012}. Accordingly, the total infinitesimal strain can be defined as follows:
\begin{equation}\label{eq:StrainTotalSMA}
    \bm{\varepsilon} = - \rho \partial_{\sigma} \mathcal{G} = \mathcal{\bm{S}} : \bm{\sigma} + \bm{\alpha} \left( T - T_0 \right) + \bm{\varepsilon}^t,
\end{equation}

\noindent incorporating the contributions from the elastic, thermal and transformation strains. And the constitutive relation for the entropy reads,
\begin{equation}
    s= - \partial_T \mathcal{G} = \frac{1}{\rho} \bm{\alpha} : \bm{\sigma} + c \ln \left( \frac{T}{T_0} \right) + s_0.
\end{equation}

The criteria determining the onset of transformation is given by,
\begin{equation}
    \dot{\xi} \Phi = 0 \, \, ,  \,\,\,\,\,\,\,\,\,\,\, \text{with} \,\,\,\,\,\, \Phi = \begin{cases}
    \Phi_{fwd}= \pi^t - Y \leq 0 & \quad \text{if } \dot{\xi}>0 \\
    \Phi_{rev}= -\pi^t - Y \leq 0 & \quad \text{if } \dot{\xi}<0
  \end{cases} \,\,\, ,
\end{equation}

\noindent where $\Phi$ is the transformation surface and $\pi^t$ is the thermodynamic driving force for transformation, work conjugate to $\xi$. The latter can be defined as:
\begin{equation}
  \pi^t = \bm{\sigma} : \bm{\Lambda}^t + \frac{1}{2} \bm{\sigma} : \left( \mathcal{S}^M - \mathcal{S}^A \right) : \bm{\sigma} + \rho \left( s_0^M - s_0^A \right) T - \rho \left( u_0^M - u_0^A \right) - f^t \, ,
\end{equation}

\noindent where the transformation direction tensor $\bm{\Lambda}^t$ and the hardening function $f^t$ take their forward or reverse form for $\dot{\xi}>0$ and $\dot{\xi}<0$, respectively.

\subsection{Variational phase field fracture}
\label{Sec:PhaseFieldTheory}

Fracture and fatigue in SMAs is predicted by using a variational phase field model \cite{Bourdin2000,Miehe2010a,Tanne2018}. Since the early work by Francfort and Marigo \cite{Francfort1998}, phase field fracture models aim at providing a variational framework for the concept of crack advance driven by the competition between toughness (surface energy density, $G_c$) and energy release rate $G$; as first proposed by Griffith \cite{Griffith1920} for elastic solids, and later extended to account for inelastic energy dissipation by Orowan \cite{Orowan1948}. In the present study, we will consider both a constant material toughness $G_c$ and the case in which the critical energy release rate is determined from its austenite and martensite counterparts, $G_c (\xi )$, using the rule of mixtures.\\ 

The discrete crack is approximated through an auxiliary field variable $\phi$, which varies between $\phi=0$, intact material, and $\phi=1$, fully cracked material. The size of the regularized crack surface is governed by the choice of $\ell$, the phase field model-inherent length scale. Thus, the fracture energy due to the creation of a crack can be approximated as:
\begin{equation}
\int_\Gamma G_c \left( \xi \right)  \, \text{d} \Gamma  \approx \int_\Omega G_c \left( \xi \right) \gamma_{\ell} \left( \phi , \, \nabla \phi \right) \, \text{d} V = \int_\Omega \frac{G_c \left( \xi \right)}{4 c_w} \left( \frac{w(\phi)}{\ell} + \ell|\nabla \phi|^2 \right) \text{d} V \, ,
\label{eq:surfaceenergy}
\end{equation}

\noindent where $\gamma_{\ell} \left( \phi , \, \nabla \phi \right)$ is the crack density function, which is itself a function of $w \left( \phi \right)$ and $c_w$. The most exploited constitutive choices of $w \left( \phi \right)$ and $c_w$ are those associated with the so-called AT1 \cite{Pham2011} and AT2 \cite{Ambrosio1991,Bourdin2000} models:
\begin{align}
  & \text{AT1:} \,\,\, w(\phi) = \phi \, , \,\,\,\, c_w=2/3 \\ 
  & \text{AT2:} \,\,\, w(\phi) = \phi^2 \, , \,\,\,\, c_w=1/2
\end{align}

The main difference between them is the fact that the AT1 model has a non-zero elastic limit. Other constitutive choices have been proposed, such as the so-called phase-field regularized cohesive zone model (PF-CZM) \cite{Wu2017}; the reader is referred to Ref. \cite{Mandal2019} for a detailed numerical comparison. To retain generality, we proceed to present our phase field formulation without making specific constitutive choices for $w \left( \phi \right)$ and $c_w$, as both the AT1 and the AT2 models will be considered in this study.\\

The total potential energy can be expressed as a function of the contributions from the mechanical, thermal and fracture terms as:
\begin{align}
\Psi = & \Psi^s \left( \bm{\varepsilon}, \xi, \phi \right) + \Psi^T  \left( T, \xi \right)+ \Psi^\phi \left( \phi, \xi \right) =   \int_\Omega \bigg\{ \left( 1 - \phi \right)^2  \psi \left( \bm{\varepsilon}, T, \xi \right) +   \nonumber \\  
& +  c \left( \xi \right) \left[ \left( T- T_0 \right) - T \ln \left( \frac{T}{T_0} \right) \right]+ \frac{G_c (\xi)}{4 c_w} \left( \frac{w(\phi)}{\ell} + \ell|\nabla \phi|^2 \right) \bigg\} \, \text{d} V \, ,
\label{Eq:Piphi}
\end{align}

\noindent where $\psi \left( \bm{\varepsilon}, T, \xi \right) $ is the strain energy density of the solid. Recall, see Eq. (\ref{eq:StrainTotalSMA}), that the strain tensor additively decomposes into an elastic part $\bm{\varepsilon}^e$, a thermal part $\bm{\varepsilon}^T=\bm{\alpha} \Delta T$ and a transformation part $\bm{\varepsilon}^t$. Accordingly, the total strain energy density $\psi$ can be expressed as
\begin{equation}\label{eq:PsiTotal}
    \psi \left(\bm{\varepsilon}, T, \xi \right) = \int_0^t \left( \bm{\sigma} : \dot{\bm{\varepsilon}}^e \right) \, \text{d}t + \int_0^t \left( \bm{\sigma} : \dot{\bm{\varepsilon}}^T \right) \text{d}t + \int_0^t \left( \bm{\sigma} : \dot{\bm{\varepsilon}}^{t} \right) \, \text{d}t \, .
\end{equation}

\noindent As elaborated below, in this work we assume that fracture is driven by the \emph{total} strain energy density.\\

Consider now the total potential energy of the solid, Eq. (\ref{Eq:Piphi}). The strong form can be readily derived by taking the first variation with respect to $\bm{\varepsilon}$, $T$ and $\phi$, and making use of Gauss' divergence theorem. Thus, the coupled field equations read, 
\begin{align}\label{eqn:strongForm}
(1-\phi)^2 \, \, \nabla \cdot \boldsymbol{\sigma}  &= \boldsymbol{0}   \hspace{3mm} \rm{in}  \hspace{3mm} \Omega \nonumber \\ 
\rho \, c \left( \xi \right) \dot{T} + \nabla \cdot \bm{q} &= 0 \hspace{3mm} \rm{in}  \hspace{3mm} \Omega \nonumber \\
G_{c} \left( \xi \right)  \left( \dfrac{\phi}{\ell}  - \ell \Delta \phi \right) - 2(1-\phi) \, \psi  &= 0 \hspace{3mm} \rm{in} \hspace{3mm} \Omega  
\end{align}

\noindent where $\bm{q}$ is the heat flux per unit area of the solid.

\subsubsection{Phase field fatigue}
\label{sec:FatigueTheory}

Now we proceed to incorporate a fatigue degradation function $f(\overline{\alpha}(t))$, extending the recent work by Carrara \textit{et al.} \cite{Carrara2020} to SMAs. The field equation corresponding to the phase field variable, (\ref{eqn:strongForm}c) is therefore given by:
\begin{equation}
    f(\overline{\alpha}(t)) \, G_{c} \left( \xi \right)  \left( \dfrac{\phi}{\ell}  - \ell \Delta \phi \right) - 2(1-\phi) \, \psi = 0 \, ,
\end{equation}

\noindent where $\overline{\alpha}$ is a cumulation of any scalar quantity which can describe the fatigue history experienced by the material. For a pseudo-time $\tau$, the cumulative history variable can be defined as \cite{Carrara2020,Alessi2018c}:
\begin{equation}\label{eq:Alpha}
    \overline{\alpha}(t)= \int_0^t H (\alpha \dot{\alpha})|\dot{\alpha}|\,\text{d}\tau,
\end{equation}

\noindent where $H(\alpha \dot{\alpha})$ is the Heaviside step function, defined as $H(\alpha \dot{\alpha})=1$ if $\alpha \dot{\alpha} \geq 0$ (loading) and $H (\alpha \dot{\alpha} ) =0$ otherwise (unloading). Accordingly, $\overline{\alpha}$ only grows during loading. It remains to define a fatigue history variable $\alpha$, representing the loading condition in the solid, and a fatigue degradation function $f(\overline{\alpha}(t))$, characterising the sensitivity of the fracture energy to the number of cycles. Regarding the former, we follow an energetic approach and define $\alpha=g(\phi)\psi$. While the degradation function $f(\overline{\alpha}(t))$ is chosen so as to vanish asymptotically,
\begin{equation}
    f(\overline{\alpha}(t)) = \begin{cases} \hspace{1cm}1 &  \text{if }\hspace{0.25cm} \overline{\alpha}(t) \leq \alpha_T \\[0.4cm] \left(\dfrac{2\alpha_T}{\overline{\alpha}(t)+\alpha_T}\right)^2 &  \text{if }\hspace{0.25cm} \overline{\alpha}(t) \leq \alpha_T \end{cases}.
\end{equation}
Here, $\alpha_T$ represents a threshold value, below which the fracture energy remains unaffected; as in Ref. \cite{TAFM2020}, we define it as:
\begin{equation}
    \alpha_T = \frac{G_c}{12 \ell}
\end{equation}

\section{Finite element implementation}
\label{Sec:NumModel}

We proceed to describe the finite element implementation of the constitutive SMA material model presented in Section \ref{sec:ConstitutiveSMA}, as well as the phase field coupled fracture/fatigue formulation described in Section \ref{Sec:PhaseFieldTheory}, including details on damage irreversibility, strain energy decomposition and solution schemes. For simplicity, the temperature will be considered to be uniform throughout our numerical experiments but the extension to non-isothermal conditions is straightforward (see, e.g., \cite{CMAME2018,Nguyen2019b,Noii2019}). The implementation is carried out in the commercial finite element package Abaqus by means of a user element (UEL) subroutine. Abaqus2Matlab is employed to pre-process the input files \cite{AES2017}.

\subsection{Implicit integration of the SMA constitutive model}

The implementation of the constitutive model follows the work by Qidwai and Lagoudas \cite{Qidwai2000}, where the evolution of the transformation strain tensor is incrementally integrated using the backward Euler method. Thus, we use the common notation of adding the subscript $n+1$ to quantities in the current time step while the subscript $n$ denotes the previous time step, e.g. for the transformation strain tensor:
\begin{equation}\label{eq:exampleN+1}
\bm{\varepsilon}^t_{n+1}=\bm{\varepsilon}^t_{n} + \Delta \bm{\varepsilon}^t_{n+1} .
\end{equation}

In addition, as the problem is solved in an iterative manner, the superscript $(k)$ is used for the iteration counter; i.e., (\ref{eq:exampleN+1}) becomes
\begin{equation}\label{eq:exampleN+1k}
 \bm{\varepsilon}^{t(k+1)}_{n+1}=\bm{\varepsilon}^{t(k)}_{n} + \Delta \bm{\varepsilon}^{t(k)}_{n+1} .   
\end{equation}

The incremental stresses are computed using a return mapping algorithm; a purely elastic trial state is followed by a transformation correction stage. Thus, assuming a uniform temperature and denoting $\mathcal{\bm{C}}$ as the elastic stiffness matrix, the elastic stress prediction is given by
\begin{equation}
    \bm{\sigma}_{n+1} = \bm{\sigma}_{n} + \mathcal{\bm{C}} \Delta \bm{\varepsilon}
\end{equation}

The return mapping algorithm is then used to enforce satisfying the transformation consistency condition $\dot{\Phi}=0$. Specifically, the convex cutting plane algorithm \cite{Ortiz1986} is used, which differs from the commonly used closest point projection algorithm as follows. Considering the flow rule (\ref{eq:flowrule}) and (\ref{eq:exampleN+1k}), the incremental transformation tensor reads,
\begin{equation}\label{eq:projectionalgorithm}
    \Delta \bm{\varepsilon}^{t(k)}_{n+1} = \left( \xi_{n+1}^{(k+1)} - \xi_n \right) \bm{\Lambda}^t \left(\bm{\sigma}_{n+1}^{(k+1)} \right) - \left( \xi_{n+1}^{(k)} - \xi_n \right) \bm{\Lambda}^t \left( \bm{\sigma}_{n+1}^{(k)} \right) \, .
\end{equation}

In the convex cutting plane algorithm the implicit dependence on the transformation direction $\bm{\Lambda}^t \left(\bm{\sigma}_{n+1}^{(k+1)} \right)$ is relaxed, and (\ref{eq:projectionalgorithm}) can be reformulated as,
\begin{equation}\label{eq:cuttingplanealgorithm}
    \Delta \bm{\varepsilon}^{t(k)}_{n+1} = \Delta \xi_{n+1}^{(k)} \bm{\Lambda}^t \left(\bm{\sigma}_{n+1}^{(k)} \right) \, .
\end{equation}

The total current strain is held constant during the iterative correction such that, considering (\ref{eq:StrainTotalSMA}), the incremental Cauchy stresses read,
\begin{equation}\label{eq:IncStressIter}
    \Delta \bm{\sigma}_{n+1}^{(k)} = - \mathcal{\bm{C}}_{n+1}^{(k)} \left( \Delta \mathcal{\bm{S}} \bm{\sigma}_{n+1}^{(k)} + \bm{\Lambda}_{n+1}^{t(k)} \right) \Delta \xi_{n+1}^{(k)} \, .
\end{equation}

In an iterative setting, the consistency condition implies,
\begin{equation}
    \Phi^{(k)}_{n+1} + \Delta \Phi^{(k)}_{n+1} = \Phi^{(k+1)}_{n+1} \approx 0 \, ,
\end{equation}

\noindent such that applying of the chain rule and inserting (\ref{eq:IncStressIter}) renders,
\begin{equation}\label{eq:LongPhiIter}
   \Phi_{n+1}^{(k)} - \partial_{\bm{\sigma}} \Phi_{n+1}^{(k)} : \mathcal{\bm{C}}^{(k)}_{n+1} \left( \Delta \mathcal{\bm{S}} \bm{\sigma}_{n+1}^{(k)} + \bm{\Lambda}_{n+1}^{t(k)} \right) \Delta \xi_{n+1}^{(k)} + \partial_\xi \Phi_{n+1}^{(k)} \Delta \xi_{n+1}^{(k)} \approx 0 \, .
\end{equation}

And finally, solving (\ref{eq:LongPhiIter}) for the correction in the martensitic volume fraction at iteration $(k)$ gives,
\begin{equation}
    \Delta \xi^{(k)}_{n+1} = \frac{- \Phi_{n+1}^{(k)}}{\partial_\xi \Phi_{n+1}^{(k)} - \partial_{\bm{\sigma}} \Phi_{n+1}^{(k)} : \mathcal{\bm{C}}^{(k)}_{n+1} \left( \Delta \mathcal{\bm{S}} \bm{\sigma}_{n+1}^{(k)} + \bm{\Lambda}_{n+1}^{t(k)} \right)} . 
\end{equation}

The transformation strain tensor can then be updated via (\ref{eq:exampleN+1k}) and (\ref{eq:cuttingplanealgorithm}). The updated values of the transformation strain and elastic stiffness are used to calculate an updated stress tensor via,
\begin{equation}
    \Delta \bm{\sigma}_{n+1}^{(k)} = - \mathcal{\bm{C}}^{(k)}_{n+1} \left( \Delta \mathcal{\bm{S}} \bm{\sigma}_{n+1}^{(k)} + \bm{\Lambda}_{n+1}^{t(k)} \right) \Delta \xi_{n+1}^{(k)} \, ,
\end{equation}

\noindent which is then used to compute the transformation function. The iterative procedure continues until $\Phi_{n+1}^{(k+1)} \approx 0$ or $\xi_{n+1}^{(k+1)}$ reaches the limiting values 0 or 1 \cite{Lagoudas2012}.\\

The last step involves computing the consistent material Jacobian $\mathcal{\bm{L}}$ \cite{Simo2006,Qidwai2000,Lagoudas2012}, which for isothermal conditions is given by,
\begin{equation}
    \text{d} \bm{\sigma} = \mathcal{\bm{L}} \text{d} \bm{\varepsilon} = \left( \mathcal{\bm{C}} + \frac{\mathcal{\bm{C}} \left( \Delta \mathcal{\bm{S}} \bm{\sigma} + \bm{\Lambda}^t \right)  \otimes \left( \mathcal{\bm{C}} \partial_{\bm{\sigma}} \Phi \right) }{\partial_\xi \Phi - \partial_{\bm{\sigma}} \Phi : \mathcal{\bm{C}} \left( \Delta \mathcal{\bm{S}} \bm{\sigma} +\bm{\Lambda}^t \right)}\right) \text{d} \bm{\varepsilon} \, . 
\end{equation}
\vspace{0.1mm}

\subsection{Addressing irreversibility and crack growth in compression}
\label{Sec:Irreversibility}

Fracture is assumed to be driven by the total strain energy density $\psi$; i.e., both the elastic $\psi^e$ and transformation $\psi^t$ strain energy densities contribute to cracking on equal footing. This choice is mainly phenomenological but has some physical background, as it appears sensible to assume a contribution from the transformation strains to the fracture process. Similar approaches have been adopted by other authors in relation to other inelastic quantities; see e.g., Ref. \cite{Borden2016,Alessi2018} for examples in the context of plasticity and ductile damage. To maintain resistance in compression and during crack closure, the elastic contribution to the strain energy density is decomposed into volumetric and deviatoric parts, following Amor \textit{et al.} \cite{Amor2009}. In this approach, the deviatoric and tensile volumetric components contribute to fracture but the compressive volumetric term does not. Thus, the elastic strain energy is decomposed into the following two terms:
\begin{align}
\psi_{+}^e &=\frac{1}{2} K \left( \xi \right) \langle tr \left( \bm{\varepsilon} \right) \rangle^2_{+} + \mu \left( \xi \right) \left( \bm{\varepsilon}' : \bm{\varepsilon}' \right) \\
\psi_{-}^e &=\frac{1}{2} K \left( \xi \right) \langle tr \left( \bm{\varepsilon} \right) \rangle^2_{-}
\end{align}

\noindent where $K$ and $\mu$ respectively denote the bulk and shear modulus, which would be dependent on $\xi$ in the transformation region. This elastic strain energy decomposition is implemented following the hybrid approach by Ambati \textit{et. al} \cite{Ambati2015}. One should note that the volumetric-deviatoric split is only applied to the elastic component of the strain energy density, such that
\begin{equation}
      \psi_{+} = \psi_{+}^e + \psi^t
\end{equation}

Secondly, a history variable field $\mathcal{H}$ is introduced to ensure damage irreversibility, $\phi_{n+1} \geq \phi_{n}$. To ensure irreversible growth of the phase field variable, the history field must satisfy the Kuhn-Tucker conditions
\begin{equation}
    \psi_{+} - \mathcal{H} \leq 0 \text{,} \hspace{7mm} \dot{\mathcal{H}} \geq 0 \text{,} \hspace{7mm} \dot{\mathcal{H}}(\psi_{+}-\mathcal{H})=0
    \centering
\end{equation}
\noindent for both loading and unloading scenarios. Thus, for a current time $t$, the history field can be defined as
\begin{equation}
     \mathcal{H} = \max_{\tau \in[0,t]}\psi_{+} ( \tau) \, . 
\end{equation}

\subsection{Finite element discretisation}
\label{Sec:FEdiscretisation}

We proceed to formulate the two-field weak form of the problem and subsequently derive the stiffness matrices and residuals applying a finite element discretisation. Thus, consider the total potential energy of the solid, Eq. (\ref{Eq:Piphi}), under isothermal conditions and in the absence of body forces and external tractions. The first variation of (\ref{Eq:Piphi}) with respect to $\bm{\varepsilon}$ and $\phi$, gives
\begin{align} \label{eq:weak1}
    \int_\Omega \left[ \left( 1 - \phi \right)^2 \bm{\sigma}: \delta \bm{\varepsilon} \right] \text{d} V &= 0 \, ,  \\
     \label{eq:weak2}
  \int_{\Omega} \left[ -2(1-\phi)\delta \phi \, \mathcal{H} +
        G_c \left( \xi \right) \left( \dfrac{\phi}{\ell} \delta \phi
        + \ell\nabla \phi \cdot \nabla \delta \phi \right) \right]  \, \mathrm{d}V &= 0   \, . 
\end{align}

Now make use of Voigt notation and consider a 3D solid. The displacement field $\bm{u}$ and the phase field $\phi$ are discretised as
\begin{equation}
    \bm{u}=\sum_{i=1}^{m} \bm{N}_{i}^{\bm{u}} \bm{u}_{i} \hspace{7mm} \text{and} \hspace{7mm} \phi=\sum_{i=1}^{m} N_{i} \phi_{i} 
    \centering
\end{equation}

\noindent where $\bm{N}_{i}$ is the shape function matrix, a diagonal matrix with $N_i$ in the diagonal terms. Here, \(N_{i}\) denotes the shape function associated with node \(i\), \(m\) is the total number of nodes per element, and $\bm{u}_{i}=\left\{u_{x}, u_{y}, u_{z} \right\}^{T}$ and \(\phi_{i}\) are the displacement and phase field values at node \(i\), respectively. Consequently, the corresponding derivatives can be discretised making use of the strain-displacement matrices $\bm{B}_{i}^{\bm{u}}$ and $\bm{B}_{i}^{\phi}$ as follows:
\begin{equation}
    \bm{\varepsilon}=\sum_{i=1}^{m} \bm{B}_{i}^{\bm{u}} \bm{u}_{i} \hspace{7mm} \text{and} \hspace{7mm} \nabla \phi=\sum_{i=1}^{m} \bm{B}_{i}^{\phi} \phi_{i} \, .
    \centering
\end{equation}
\noindent Here \(\bm{\varepsilon}=\left\{\varepsilon_{xx}, \varepsilon_{yy}, \varepsilon_{zz},  \gamma_{xy}, \gamma_{xz}, \gamma_{yz} \right\}^{T}\), with $\gamma$ being the engineering strain, such that $\gamma_{xy}=2 \varepsilon_{xy}$. \\

We proceed to formulate the residuals and the stiffness matrices considering this finite element discretisation and the fact that (\ref{eq:weak1})-(\ref{eq:weak2}) must hold for arbitrary values of \(\delta \bm{u}\) and \(\delta \phi\). The associated discrete equations can be formulated as the following residuals with respect to the displacement field and the phase field variable, respectively,
\begin{align}
    \bm{r}_{i}^{\bm{u}} = \int_{\Omega} \left\{  \left[(1-\phi)^{2} + \kappa \right] {(\bm{B}_{i}^{\bm{u}})}^{T} \bm{\sigma} \right\} \, \text{d}V & \\ 
    r_{i}^{\phi}= \int_{\Omega} \left\{ -2(1-\phi) N_{i} \, \mathcal{H} +
    G_{c} \left( \xi \right) \left[ \frac{\phi}{\ell} N_{i} 
    + \ell {(\bm{B}_{i}^{\phi})}^{T} \nabla \phi \right] \right\} \, \text{d}V  &  
\end{align}

\noindent where $\kappa$ is a numerical parameter introduced to keep the system of equations well-conditioned. A value of $\kappa= 1 \times 10^{-7}$ is adopted throughout this work. Finally, the tangent stiffness matrices can be readily computed by taking the first derivative of the residual vectors, rendering
\begin{align}
  \bm{K}_{ij}^{\bm{u} \bm{u}} = \frac{\partial \bm{r}_{i}^{\bm{u}} }{\partial \bm{u}_{j} } = 
        \int_{\Omega} \left\{ \left[ (1-\phi)^2+\kappa \right] {(\bm{B}_{i}^{\bm{u}})}^{T} \mathcal{\bm{L}} \, \bm{B}_{j}^{\bm{u}} \right\} \, \text{d}V & \\
        \bm{K}_{ij}^{\phi \phi} = \frac{\partial r_{i}^{\phi} }{ \partial \phi_{j} } =  \int_{\Omega} \left\{ \left( 2 \mathcal{H} + \frac{G_{c} \left( \xi \right)}{\ell} \right) N_{i} N_{j} + G_{c} \left( \xi \right) \ell \, {(\bm{B}_{i}^{\phi})}^{T} (\bm{B}_{j}^{\phi}) \right\} \, \text{d}V &
\end{align}
\vspace{0.1mm}

\subsection{Solution schemes}
\label{Sec:StaggeredSolutionScheme}

A global iterative scheme is adopted to obtain the displacement $\bm{u}$ and phase field $\phi$ solutions for which $\bm{r}^{\bm{u}}=\bm{0}$ and $\textbf{r}^{\phi}=\bm{0}$;
\begin{equation}\label{Eq:GlobalElementSystem}
    {\begin{Bmatrix}
        \textbf{u}\\[0.3em] \bm{\phi}
    \end{Bmatrix}}_{t+\Delta t} = 
    {\begin{Bmatrix}
        \textbf{u}\\[0.3em] \bm{\phi}
    \end{Bmatrix}}_{t} -
    {\begin{bmatrix}
        \mathbf{K}^{\mathbf{u}\mathbf{u}}+\mathbf{M} & 0 \\[0.3em] 
        0 & \mathbf{K}^{\phi\phi}
    \end{bmatrix}}_{t}^{-1}
    {\begin{Bmatrix}
        \textbf{r}^{\textbf{u}}\\[0.3em] \textbf{r}^{\phi}
    \end{Bmatrix}}_{t} \, .
    \centering
\end{equation}

We develop a numerical implementation that can accommodate both staggered \cite{Miehe2010a} and monolithic quasi-Newton schemes \cite{Wu2020a,TAFM2020}. As shown by Wu \textit{et al.} \cite{Wu2020a} for the PF-CZM model and by Kristensen and Mart\'{\i}nez-Pa\~neda \cite{TAFM2020} for the AT2 model, the use of quasi-Newton methods such as the Broyden–Fletcher–Goldfarb–Shanno (BFGS) algorithm enables a robust implementation of monolithic schemes that retain unconditional stability, speeding up calculations by several orders of magnitude. In this work, this is particularly relevant for the analysis of fatigue, as otherwise calculations would be prohibitive \cite{TAFM2020}. The reader is referred to Refs. \cite{Miehe2010a} and \cite{TAFM2020} for details on the implementation of staggered and monolithic quasi-Newton solution schemes, respectively. 

\section{Results}
\label{Sec:FEMresults}

We proceed to demonstrate the potential of the computational modelling framework by addressing a number of case studies of particular interest. First, in Section \ref{sec:BoundaryLayer}, a boundary layer model is used to gain insight into the fracture behaviour of SMAs by investigating stationary and propagating cracks. Secondly, in Section \ref{sec:SquarePlate}, we proceed to model mode I fracture in a square plate, a paradigmatic phase field fracture benchmark. Mixed-mode conditions and crack coalescence is then investigated using an asymmetric double-notched bar in Section \ref{sec:AsymmetricNotched}. Finally, we conduct a 3D large scale analysis of fatigue failure of a biomedical stent in Section \ref{sec:FatigueStent}.\\

\begin{figure}[H]
    \centering
    \includegraphics[width=0.8\linewidth]{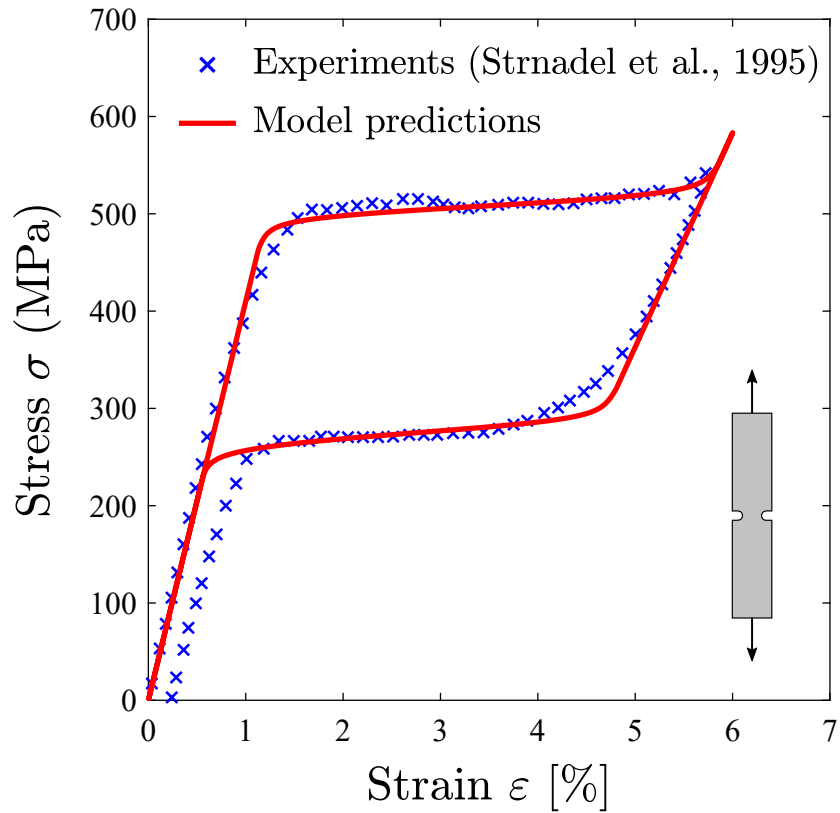}
    \caption{Uniaxial tensile stress-strain response of NiTi showing the validation of the model against the experimental data by Strnadel \textit{et al.} \cite{Strnadel1995}.}
    \label{fig:Uniaxial}
\end{figure}

Our numerical experiments are conducted on an equiatomic nitinol SMA, following the experimental data provided in Refs. \cite{Strnadel1995,Lagoudas2008}. The phase diagram transformation surface slopes for martensite ($C_M$) and austenite ($C_A$) are given at a reference stress of $\sigma^*=300$ MPa. A uniform temperature of $T=320$ K is generally adopted, following the experiments, but its influence will be investigated. In addition, we assume a uniform transformation strain, such that in (\ref{eq:Hmaxmin}), $H=H_{min}=H_{max}$. Regarding the material toughness, a value of 22.5 kJ/m$^2$ is adopted from the range of reported data for NiTi \cite{Haghgouyan2018}, unless otherwise stated. However, one should note that this choice is based on the austenite fracture resistance, as a consequence of the assumption of a small scale transformation zone. The implications of this assumption will be discussed and results compared to the case of a martensite volume fraction-dependent critical energy release rate, $G_c \left( \xi \right)$. In regards to the constitutive choices inherent to the phase field model, the conventional AT2 model is generally adopted, although calculations are also conducted with the AT1 model for comparative purposes. The comparison between the calibrated model predictions and the experimental data from the uniaxial tension tests by Strnadel \textit{et al.} \cite{Strnadel1995} is shown in Fig. \ref{fig:Uniaxial}. The smooth hardening capabilities of the model enable attaining a very good agreement with the experiments.

\begin{table}[H]
\caption{Selected material parameters used in the numerical experiments, following the measurements by Strnadel \textit{et al.} \cite{Strnadel1995} on an equiatomic nitinol SMA.}
\raggedleft
\hspace*{-3cm} 
\begin{tabular}{l l} 
\thickhline
Parameter & Magnitude\\
\thickhline
Elastic properties\\
\hline
Austenite's Young's modulus, $E_A$ (MPa) & 41000 \\
Martensite's Young's modulus, $E_M$ (MPa) & 22000\\
Austenite's Poisson's ratio, $\nu_A$ & 0.33 \\
Martensite's Poisson's ratio, $\nu_M$ & 0.33  \\
\hline
Phase diagram properties\\
\hline
Martensite start temperature, $M_s$ (K) & 239 \\
Martensite end temperature, $M_f$ (K) & 221 \\
Austenite start temperature, $A_s$ (K) & 266 \\
Austenite end temperature, $A_f$ (K) & 282 \\
$\sigma$ vs $T$ slope (loading), $C_M|_{\sigma=300 \, \text{MPa}}$ (MPa/K) & 5.5 \\
$\sigma$ vs $T$ slope (unloading), $C_A|_{\sigma=300 \, \text{MPa}}$ (MPa/K) & 5.5\\
\hline
Other\\
\hline
Transformation strain $H$ & 0.0335 \\
Material toughness $G_c$ (kJ/m$^2$) & 22.5 \\
Smooth hardening properties $n_1$, $n_2$, $n_3$, $n_4$ & 0.15, 0.17, 0.25, 0.15 \\
Temperature $T$ (K) & 320 \\
\thickhline
\end{tabular}
\label{tab:tab1length}
\end{table}

\subsection{Boundary layer formulation}
\label{sec:BoundaryLayer}

\begin{figure}[H]
    \centering
    \includegraphics[width=1.1\linewidth]{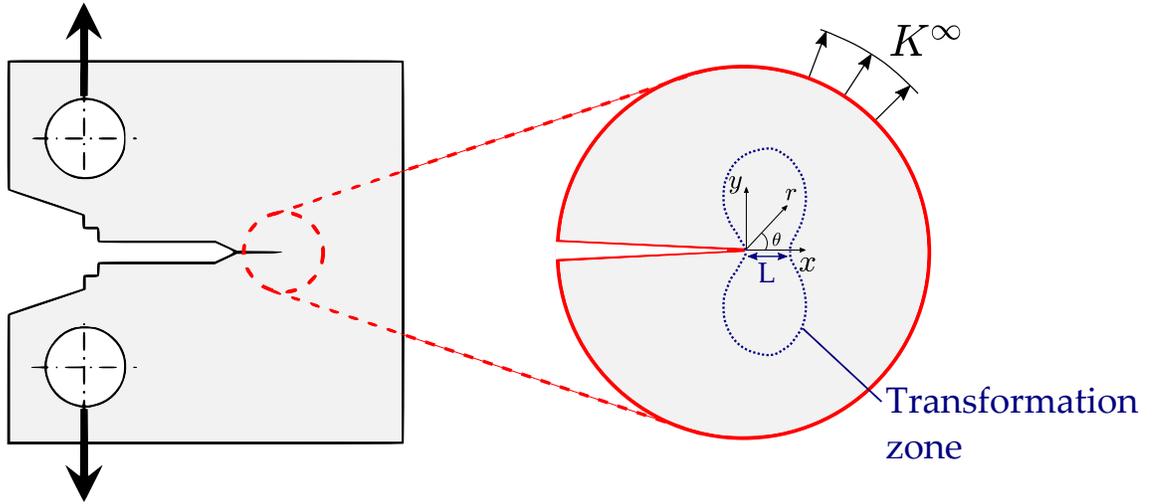}
    \caption{Boundary layer concept, illustrated on a Compact Tension specimen.}
    \label{fig:BoundaryLayer}
\end{figure}

The concept of a boundary layer formulation is illustrated in Fig. \ref{fig:BoundaryLayer}. For a cracked solid, the crack tip stress state is characterised by the stress intensity factor; $K_I$, assuming mode I conditions. The Williams \cite{Williams1957} solution for a linear elastic solid can be used to relate the displacement field to the magnitude of $K_I$. Considering a polar coordinate system $(r, \theta)$ and a Cartesian coordinate system $(x, y)$ centred at the crack tip, with the crack plane along the negative $x$-axis, the displacement solution reads:
\begin{equation}\label{eq:Williams1}
u_i = \frac{K_I}{E_A} r^{1/2} f_i \left( \theta, \nu_A \right),
\end{equation}
\noindent where the subscript index $i$ equals $x$ or $y$, and the functions $f_i \left( \theta, \nu \right)$ are given by
\begin{equation}\label{eq:Williams2}
f_{x} = \frac{1+\nu_A}{\sqrt{2 \pi}} \left(3 - 4 \nu_A - \cos \theta \right) \, \cos \left(\frac{\theta}{2} \right)
\end{equation}
\noindent and
\begin{equation}\label{eq:Williams3}
f_{y} = \frac{1+\nu_A}{\sqrt{2 \pi}} \left(3 - 4 \nu_A - \cos \theta \right) \, \sin \left(\frac{\theta}{2} \right).
\end{equation}

Thus, the nodal displacements in the outer boundary of the finite element model can be prescribed to evaluate the crack tip behaviour at a given value of $K_I$. However, one should note that this is under the assumption that the inelastic region is small, generally referred to as small transformation zone conditions; analogous to the so-called small scale yielding conditions in elastic-plastic materials. Accordingly, Eqs. (\ref{eq:Williams1})-(\ref{eq:Williams2}) make use of the elastic constants for the austenitic phase.

\subsubsection{Stationary crack tip fields}

\begin{figure}[H]
  \makebox[\textwidth][c]{\includegraphics[width=0.95\textwidth]{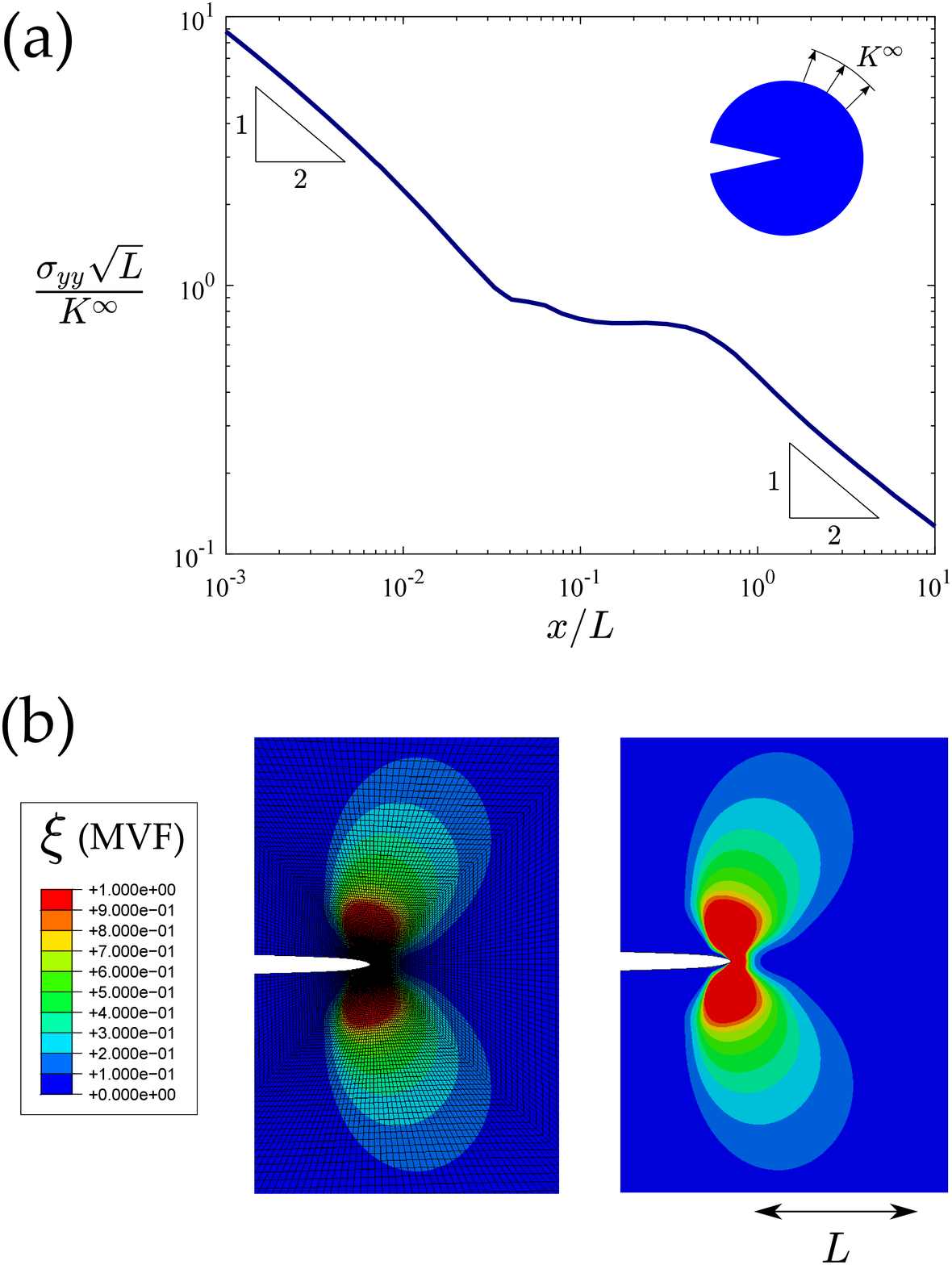}}%
  \caption{Crack tip fields ahead of a stationary crack: (a) normalised crack tip stresses versus distance to the crack tip, and (b) contours of the martensite volume fraction $\zeta$. $L$ denotes the characteristic length scale associated with $K^\infty$, as given by Eq. (\ref{eq:Lreference}).}
  \label{fig:Stationary1}
\end{figure}

The analysis of stationary cracks provides insight into the fracture behaviour of SMAs and facilitates interpretation of the phase field fracture results. We adopt a boundary layer formulation, as described in Fig. \ref{fig:BoundaryLayer}, and take advantage of symmetry to model only the upper half of the circular domain. After a mesh sensitivity study, a total of 14,183 quadrilateral quadratic elements with reduced integration are used. The mesh is refined close to the crack tip, as shown in Fig. \ref{fig:Stationary1}b. The loading is applied by prescribing a remote $K_I$ field following Eqs. (\ref{eq:Williams1})-(\ref{eq:Williams2}). Accordingly, a reference length scale can be defined as \cite{Freed2007}:
\begin{equation}\label{eq:Lreference}
    L = \frac{1}{2 \pi} \left[ \frac{K_I}{C_M (T - M_f )} \right]^2 \, .
\end{equation}

The finite element results obtained for a stationary crack in the reference NiTi material of Table \ref{tab:tab1length} are shown in Fig. \ref{fig:Stationary1}. Consider first Fig. \ref{fig:Stationary1}a, where the normalised crack tip stress distribution is shown as a function of the normalised distance ahead of the crack tip. In agreement with expectations, a stress-induced transformation region develops near the crack tip of SMAs. Three distinct domains can be observed. Adjacent to the crack tip, an inner $K_I$ field is observed within the martensitic region, where crack tip stresses exhibit the $r^{-1/2}$ linear elastic singularity. For distances larger than roughly 0.03$L$ from the crack tip, a transformation region exists, where the stresses are much less singular and the martensitic volume fraction is between 0 and 1, see Fig. \ref{fig:Stationary1}b. Farther away from the crack tip, the $r^{-1/2}$ linear elastic singularity is again recovered, indicating the presence of an outer $K_I$ field in the purely austenitic region ($\xi=1$). For clarity, this outer $K_I$ field associated with the austenitic phase is here frequently denoted as $K^\infty$. Depending on the material properties, the size of the sample and the temperature, the outer $K^\infty$ regime might be very small, which would complicate fracture mechanics testing - see Ref. \cite{Haghgouyan2019} for a discussion. Also, the consideration of plastic yielding will predictably introduce an additional crack tip region, adjacent to the crack tip and within the inner elastic domain. Fig. \ref{fig:Stationary1}b shows the shape of the transformation zone, with blue and red colours respectively denoting the martensitic and austenitic phases. It follows immediately from the existence of this stress-induced transformation region that the $J$-integral becomes path-dependent; see, e.g. Ref. \cite{IJMMD2015} for a discussion on the path-dependence of $J$ in inhomogeneous materials.

\subsubsection{Crack growth resistance curves (R-curves)} 

We proceed to model crack advance using the phase field fracture formulation described in Section \ref{Sec:PhaseFieldTheory}. As in the stationary crack analysis, a boundary layer model is used. In this case, the refined region of the finite element mesh extends over the entire crack propagation domain, where in all calculations the characteristic element size is at least seven times smaller than the phase field length scale, following Ref. \cite{CMAME2018}. Approximately 47,200 quadrilateral linear elements are used. Crack growth resistance curves (R-curves) are predicted by computing the crack extension $\Delta a$ as a function of the remote load, as characterised by the remote stress intensity factor $K_I$. The remote load is normalised by a reference stress intensity factor, given by the following relationship:
\begin{equation}\label{eq:K0}
    K_0= \sqrt{\frac{E_A G_c}{(1-\nu_A^2)}} \, ,
\end{equation}

\noindent while the crack extension is normalised by a critical length $L_c$ \cite{Freed2007}, associated with the material toughness,
\begin{equation}
    L_c  =  \frac{(1-2\nu_A)^2}{2 \pi (1-\nu_A^2)} \frac{E_A G_c}{\left[ C_M (T - M_s ) \right]^2} \, . 
\end{equation}

\begin{figure}[H]
  \makebox[\textwidth][c]{\includegraphics[width=1.4\textwidth]{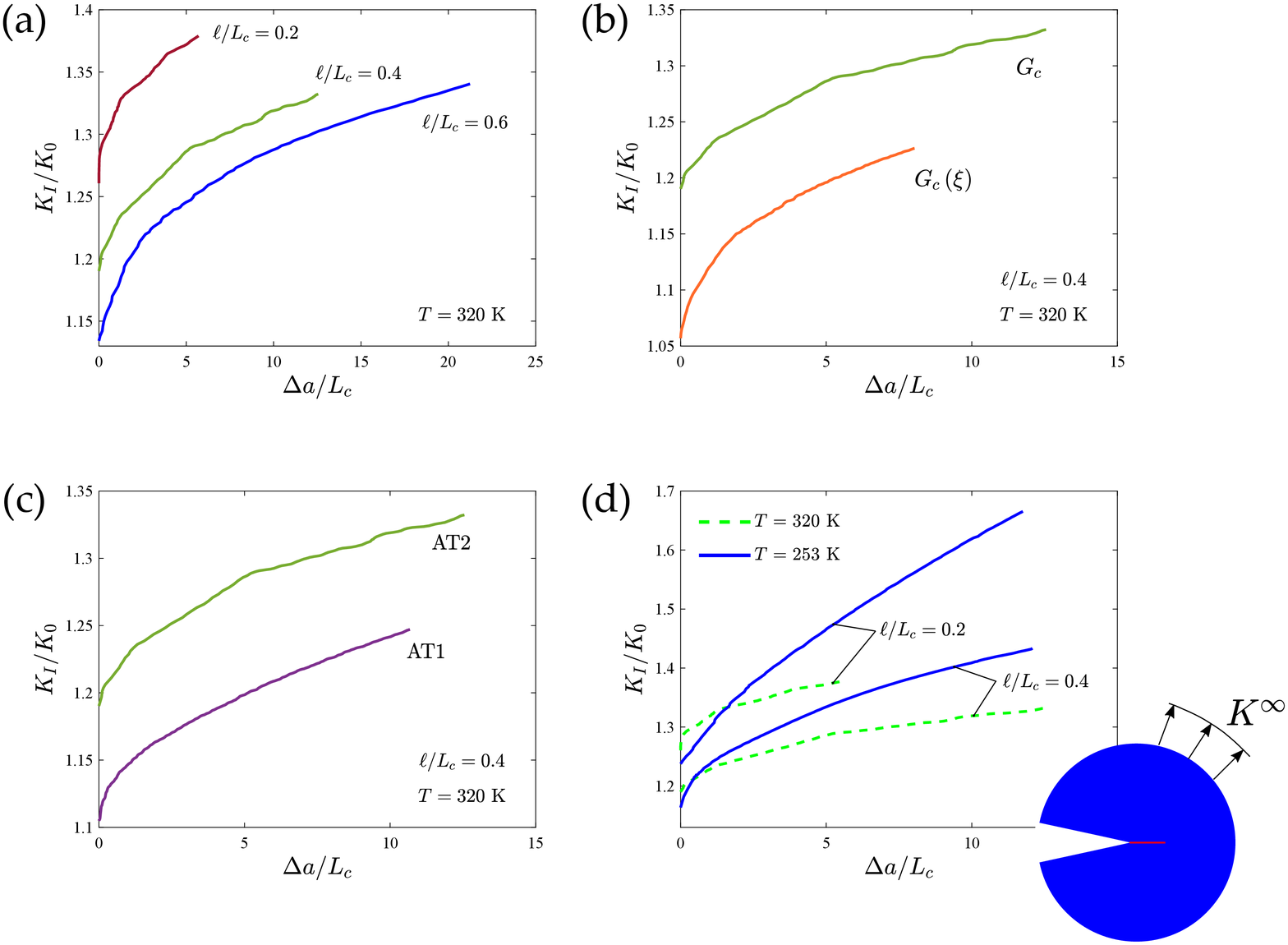}}%
  \caption{Crack growth resistance in SMAs, influence of: (a) the phase field length scale, (b) the toughness homogenisation, (c) the phase field crack density function, and (d) the temperature. Material properties as defined in Table \ref{tab:tab1length}.}
  \label{fig:Rcurve}
\end{figure}

The results, shown in Fig. \ref{fig:Rcurve}, examine the influence of: (a) the phase field length scale, (b) the toughness definition, (c) the crack density function and (d) the temperature. In all cases, the model predicts a toughening effect (rising R-curve) associated with energy dissipation due to phase transformation, as observed in the experiments. 

Consider first Fig. \ref{fig:Rcurve}a, where the results reveal a rising R-curve with decreasing $\ell/L_c$ ratio. The trends observed can be explained as follows; recall (see, e.g., \cite{CMAME2018,Tanne2018}) that the phase field length scale is related to the material tensile strength as,
\begin{equation}
   \hat{\sigma} = \frac{9}{16} \sqrt{\frac{E G_c}{3 \ell}} \, .
\end{equation}

Consequently, smaller values of $\ell$ lead to higher strengths and this results in greater inelastic dissipation, as it has been observed using cohesive zone models in SMAs \cite{Freed2007} and in elastic-plastic materials \cite{Tvergaard1992,JMPS2019}. However, note that in previous crack growth analyses in SMAs, the fracture energy was assumed to be constant. As shown in Fig. \ref{fig:Rcurve}b, we proceed to define $G_c$ as a function of the martensite volume fraction using the rule of mixtures,
\begin{equation}\label{eq:rulemixtures}
     G_c \left( \xi \right) = \left( 1 - \xi \right) G_c^A  + \xi G_c^M,
\end{equation}
\noindent and evaluate its influence. Following Ref. \cite{Haghgouyan2019}, we take the toughness of the martensite phase ($G_c^M$) to be 20\% smaller than that of the austenitic one, provided in Table \ref{tab:tab1length}. The results show that cracking initiates at a lower load level and this results in lesser dissipation. This is not surprising given the smaller magnitude of $G^M_c$ and the fact that cracking initiates in the martensitic region. Note that $K_0$ has been defined relative to the remote $K_I$, using the elastic constants for the austenitic phase, see (\ref{eq:K0}); this results in a higher prediction of the initiation load, which occurs when $G=G_c$ is met locally. As shown in Ref. \cite{JMPS2020} for homogeneous elastic-plastic materials, the onset of crack growth in the presence of a large initial crack is based only on energy considerations; i.e., it occurs at $K_I=K_0$ and it is insensitive to the value of $\ell$. In the SMA case, cracking initiates at $K_I > K_0$ due to: (i) the differences between the inner and outer $K$-fields, as discussed in the stationary crack analysis, and (ii) the $\xi$-dependence of $G_c$. The choice of a $\xi$-dependent fracture energy, not considered so far (see, e.g., \cite{Freed2007,Baxevanis2014,Karimi2019}), appears to be a sensible one. Experiments show significant differences between the toughness of purely austenitic and purely martensitic samples \cite{Haghgouyan2019}, and the results obtained here (Fig. \ref{fig:Rcurve}b) reveal non-negligible differences in the predicted crack growth resistance behaviour relative to the choice of a constant fracture energy.\\

Next, the influence of the constitutive choice for the phase field crack density function is assessed in Fig. \ref{fig:Rcurve}c. The results reveal that cracking initiates earlier in the AT1 model and leads to less dissipation, relative to the AT2 case. Finally, the role of temperature is quantified in Fig. \ref{fig:Rcurve}d. This is of interest because the reference temperature (320 K) is above the austenitic end temperature ($A_f$), which should lead to a full recovery in the wake of the crack. This is not the case for 253 K, which is below the austenitic start temperature ($A_s$), implying that no reverse transformation takes place upon unloading. The unloading response has proven to have an important effect in elastic-plastic materials, revealing big differences between isotropic and kinematic hardening laws \cite{JAM2018,EFM2019}. In the case of SMAs, a higher degree of dissipation is observed in the cases with smaller temperature, where the magnitude of the austenite to martensite transformation stresses is smaller.

\subsection{Fracture of a square plate with a crack}
\label{sec:SquarePlate}

\begin{figure}[H]
    \centering
    \includegraphics[width=0.6\linewidth]{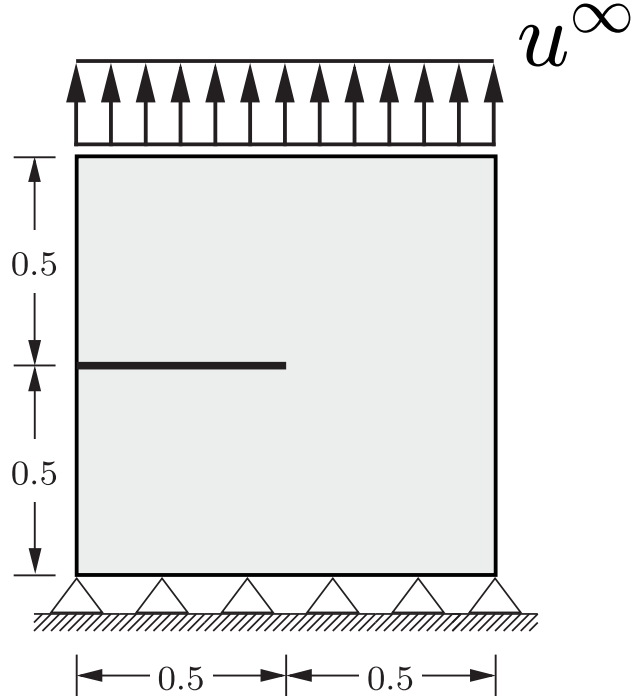}
    \caption{Cracked square plate: dimensions (in mm) and loading configuration.}
    \label{fig:CrackedSquare}
\end{figure}

We proceed to model the failure of a cracked square plate subjected to tension, a paradigmatic benchmark in phase field fracture \cite{Miehe2010a,Ambati2015,CMAME2018,TAFM2020}. The specimen has an initial horizontal crack going from the left side to the center of the specimen, both vertical and horizontal displacements are restricted in the bottom boundary, and we load the plate by prescribing the vertical displacement in the upper edge, $u^\infty$. The geometric setup, dimensions (in mm) and boundary conditions are given in Fig. \ref{fig:CrackedSquare}. The constitutive behaviour of the material is characterised by the parameters listed in Table \ref{tab:tab1length} while the phase field model uses $G_c=4.1$ kJ/m$^2$ and $\ell=0.0075$ mm. The characteristic element size is at least 7 times smaller than $\ell$ along the extended crack plane. A total of 45,571 quadrilateral linear elements with full integration are used.

\begin{figure}[H]
\centering
\noindent\makebox[\textwidth]{%
\includegraphics[scale=1]{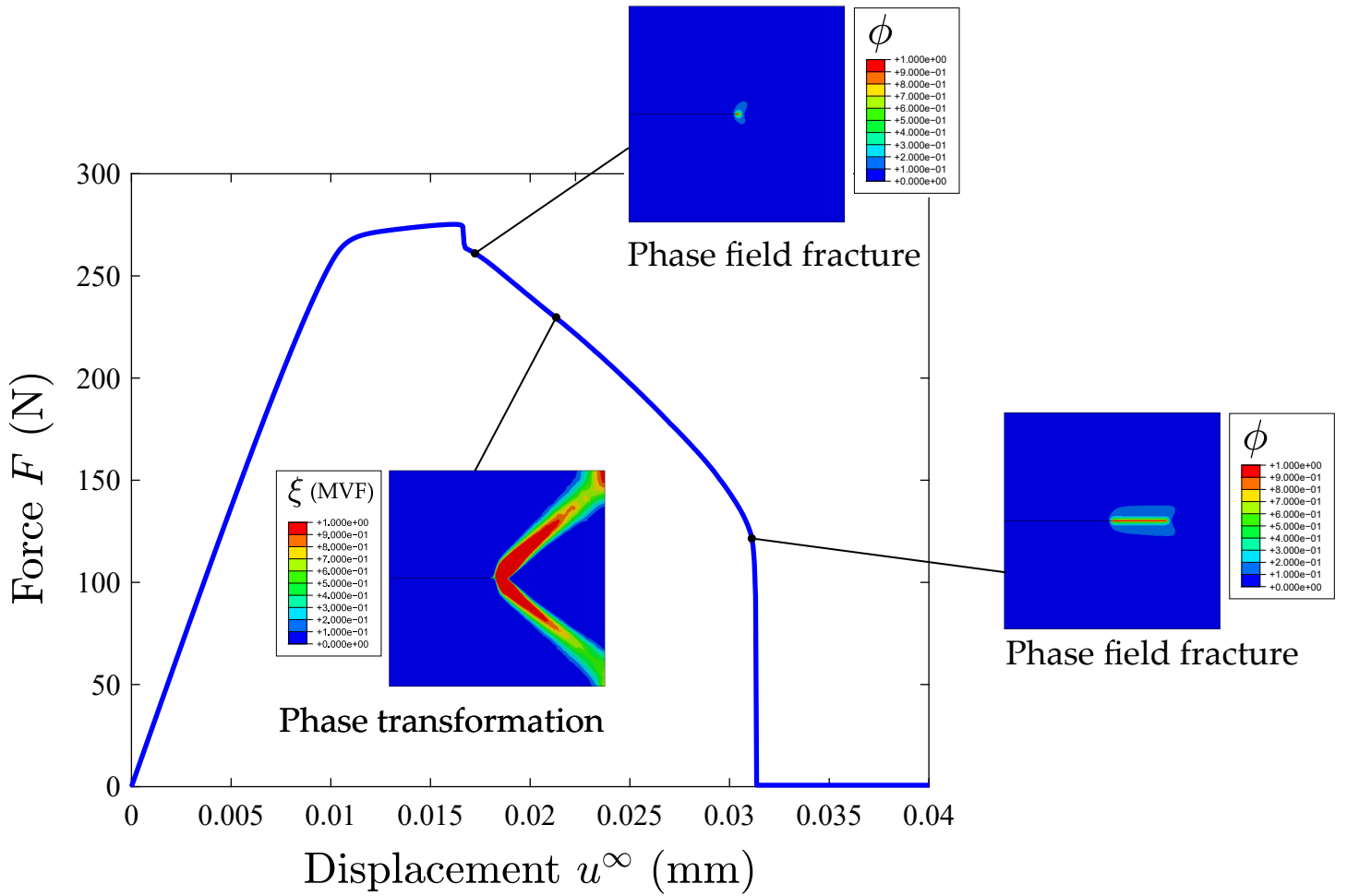}}
\caption{Cracked square plate: force versus displacement response with contours of martensite volume fraction $\xi$ and phase field fracture parameter $\phi$.}
\label{fig:CrackedSquareR}
\end{figure}

The results obtained are shown in Fig. \ref{fig:CrackedSquareR} in terms of the force versus displacement response, with contours of martensite volume fraction $\xi$ and phase field fracture parameter $\phi$ at different stages embedded in the figure. Differences from the response commonly observed in this classic benchmark are notable. In a linear elastic homogeneous material the plate fails in an unstable manner, with the force versus displacement curve exhibiting a very large drop immediately after reaching the peak load - see, e.g. \cite{Miehe2010a,TAFM2020}. Contrarily, in the SMA sample a significant toughening effect is observed; there is an initial drop in the load associated with the first instance of crack growth but then the crack progresses in a stable manner until complete failure of the sample. The toughening effect observed is undoubtedly related to the energy dissipated due to transformation. As shown in Fig. \ref{fig:CrackedSquareR}, a large transformation zone develops in the sample, exhibiting a shear banding-like behaviour that resembles that observed in elastic-plastic materials \cite{Molnar2020}. However, one should note that, for the material properties here considered, the crack still follows the mode I fracture path. 

\subsection{Plane strain tension of an asymmetric double-notched specimen}
\label{sec:AsymmetricNotched}

\begin{figure}[H]
    \centering
    \includegraphics[width=0.4\linewidth]{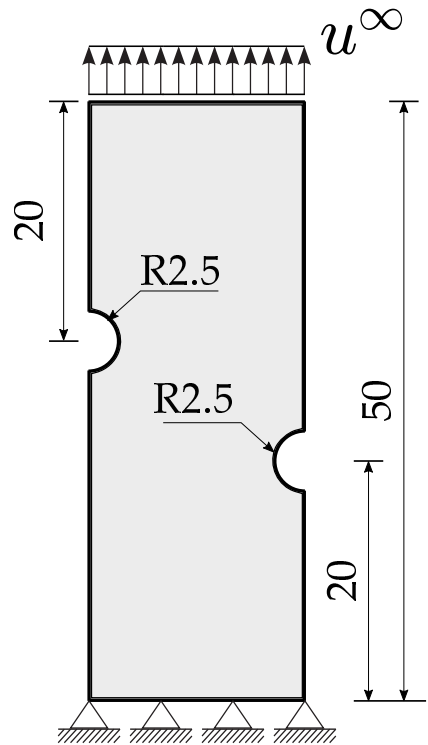}
    \caption{Plane strain tension of an asymmetric double-notched specimen: dimensions (in mm) and loading configuration.}
    \label{fig:AsymmetricDoubleNotch}
\end{figure}

An asymmetrically notched plane strain bar is investigated to model mixed-mode fracture and crack coalescence. The double-notched bar, depicted in Fig. \ref{fig:AsymmetricDoubleNotch}, is clamped at the bottom end ($u_x=u_y=0$) and subjected to a vertical displacement $u^\infty$ at the top edge. Two circular notches of radii 2.5 mm have been geometrically introduced. The bar is assumed to be made of the equiatomic nitinol SMA whose properties and model parameters are listed in Table \ref{tab:tab1length}. The phase field length scale equals 0.2 mm and the finite element mesh is chosen accordingly, with the characteristic element length in the region between the two notches being on the order of 0.05 mm. A total of 23,622 quadrilateral plane strain linear elements are used to discretise the model.

\begin{figure}[H]
\centering
\noindent\makebox[\textwidth]{%
\includegraphics[scale=0.25]{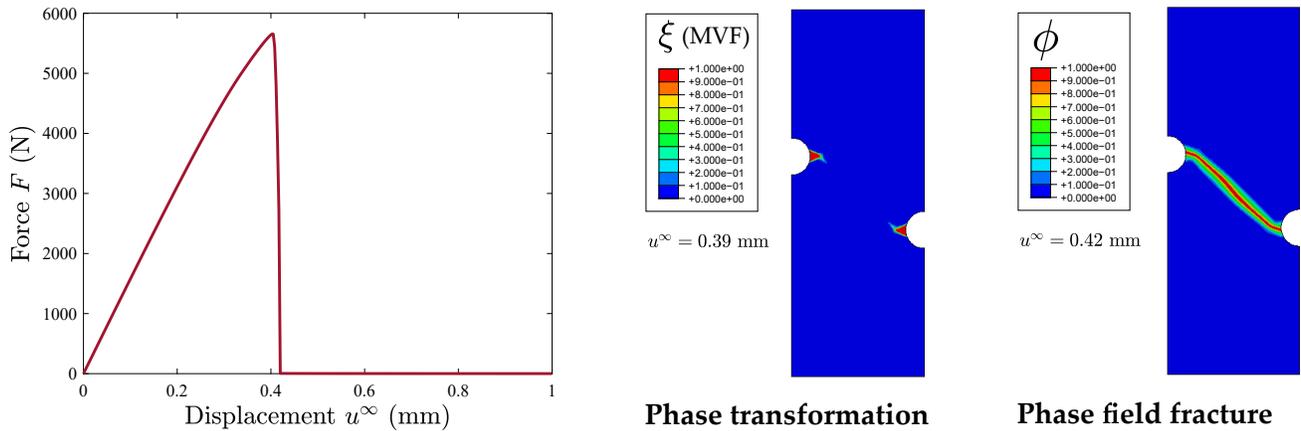}}
\caption{Plane strain tension of an asymmetric double-notched specimen: force versus displacement response. The figure includes contours of martensite volume fraction $\xi$, just before failure, and phase field fracture parameter $\phi$, after the unstable crack growth event.}
\label{fig:AsymmetricDoubleNotchResults}
\end{figure}

The finite element results computed are shown in Fig. \ref{fig:AsymmetricDoubleNotchResults}, including the force versus displacement response as well as contours of phase transformation and phase field fracture parameter. Contrarily to the response observed in the previous case study, Section \ref{sec:SquarePlate}, a sharp drop in the force versus displacement curve is observed, indicative of brittle fracture with little inelastic dissipation. This can be rationalised by observing the phase transformation contour just before failure, at $u^\infty=0.39$ mm; as shown in Fig. \ref{fig:AsymmetricDoubleNotchResults}, the inelastic region is confined to a small area in the close vicinity of the tips of the notches. No inelastic shear bands are observed such that as soon as cracking initiates at the notch tips, the cracks coalescence and unstable cracking is observed. We note that this finding is specific to the boundary value considered, a parametric analysis is needed to characterise the interplay between the different scales at play (notch radius, phase field length scale, sample dimensions) and its implications on fracture stability due to inelastic dissipation.

\subsection{Fatigue failure of a NiTi stent}
\label{sec:FatigueStent}

The capabilities of the modelling framework in capturing fatigue crack growth are demonstrated by simulating cyclic damage in a stent, a paradigmatic application of shape memory materials \cite{Reese2010,Frischkorn2015,Auricchio2015}. This case study also serves to showcase the computational efficiency of the framework and its applicability to the modelling of computationally demanding large-scale 3D boundary value problems.\\

\begin{figure}[H]
\centering
\noindent\makebox[\textwidth]{%
\includegraphics[scale=0.11]{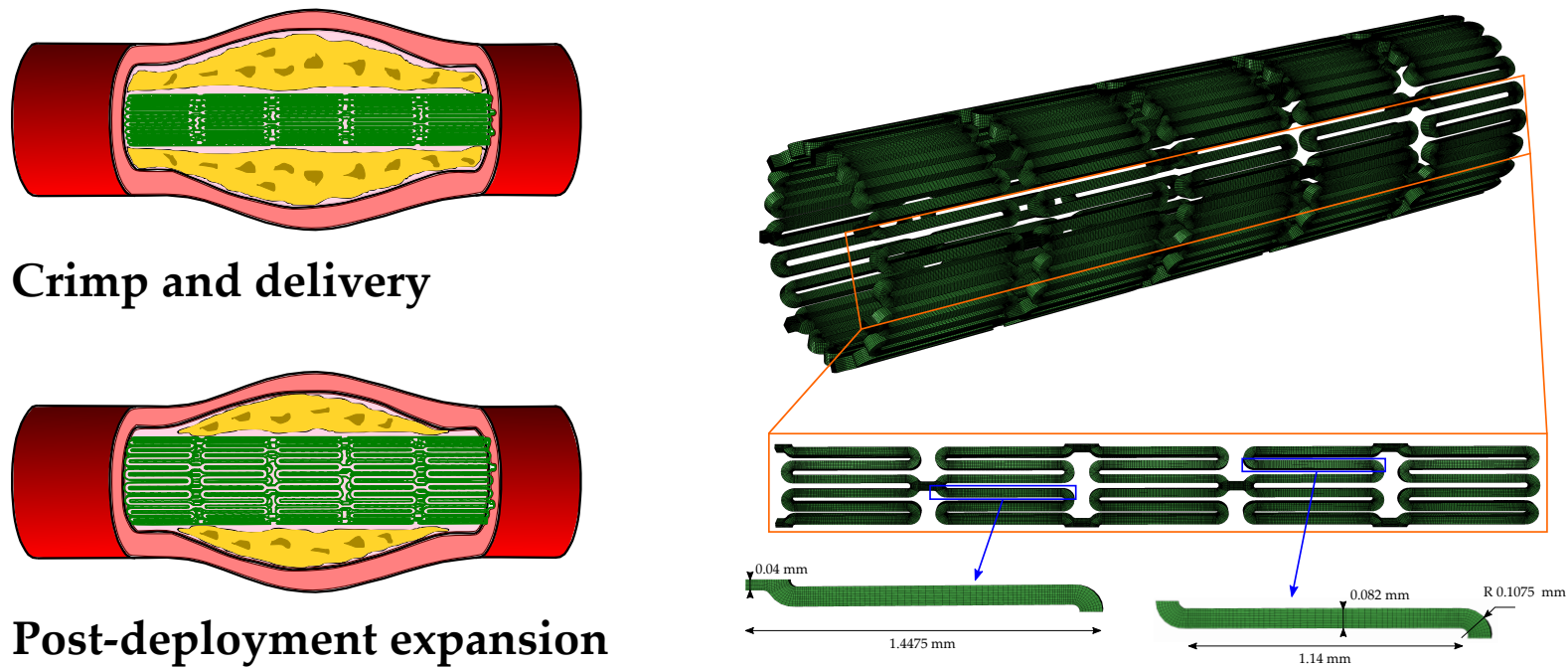}}
\caption{Sketch of the functionality of a NiTi stent, along with the geometry and mesh of the finite element model. The struts have a thickness of 0.1 mm and the stent has an inner radius of 0.8465 mm. }
\label{fig:StentSketch}
\end{figure}

Stents are small cylindrical tubes that are placed into blood vessels, arteries, or other ducts to hold the structure open. Often, their role is to counteract the effects of vascular diseases that are associated with plaque blockages that hinder fluid flow, see Fig. \ref{fig:StentSketch}. To deploy the stent, it is usually crimped, placed in a delivery system (e.g., catheter), and finally expanded in-vivo to widen the duct. Nitinol is a popular material choice in stent manufacturing due to its biocompatibility and capacity to expand by recovering its elastic deformation after the constraining delivery system has been removed (superelasticity). However, fatigue resistance is often the limiting design criterion as the stent is subjected to repeated contraction and expansion during the systolic and diastolic cycles. Current fatigue design models for stents commonly use Goodman's and other empirical methods to estimate the number of cycles to failure by extrapolating from the stress/strain state of the first cycle. We aim here at providing a more mechanistic approach using a phase field fatigue framework that can predict features such as S-N curves or the Paris law as a natural outcome of the model \cite{Carrara2020}.\\

We assume that the stent has been manufactured using the equiatomic NiTi whose material properties are listed in Table \ref{tab:tab1length}. Taking advantage of symmetry along the longitudinal direction, half of the stent is modelled. The stent is subjected to different expansion and compressive pressures by prescribing a radial displacement in the outer surface, $u_r$. First, there is a crimp phase where the simulation emulates the compression of the stent inside of the capsule prior to delivery; a radial displacement of $u_r=-0.16$ mm is applied. Secondly, the stage of deployment is reproduced by allowing the stent to expand up to $u_r=-0.02$ mm, where further expansion is limited by the surrounding duct. Once deployed, the stent is subjected to cyclic loads of amplitude $\Delta u_r=-0.04$ mm that simulate the compression and expansion pressures experienced due to the systolic and diastolic cycles. The damage response will be governed by the phase field fatigue model described in Section \ref{sec:FatigueTheory}, with a phase field length scale of $\ell=0.02$ mm and using the monolithic quasi-Newton solution scheme. The magnitude of $\ell$ is at least 2.5 times larger than the characteristic element length. Eight-node brick elements with full integration are used to discretise the geometry, with the model containing more than 7 million degrees-of-freedom (DOFs). Calculations are run in parallel, using 16 cores, with each load increment taking approximately 2 minutes.

\begin{figure}[H]
\centering
\noindent\makebox[\textwidth]{%
\includegraphics[scale=0.06]{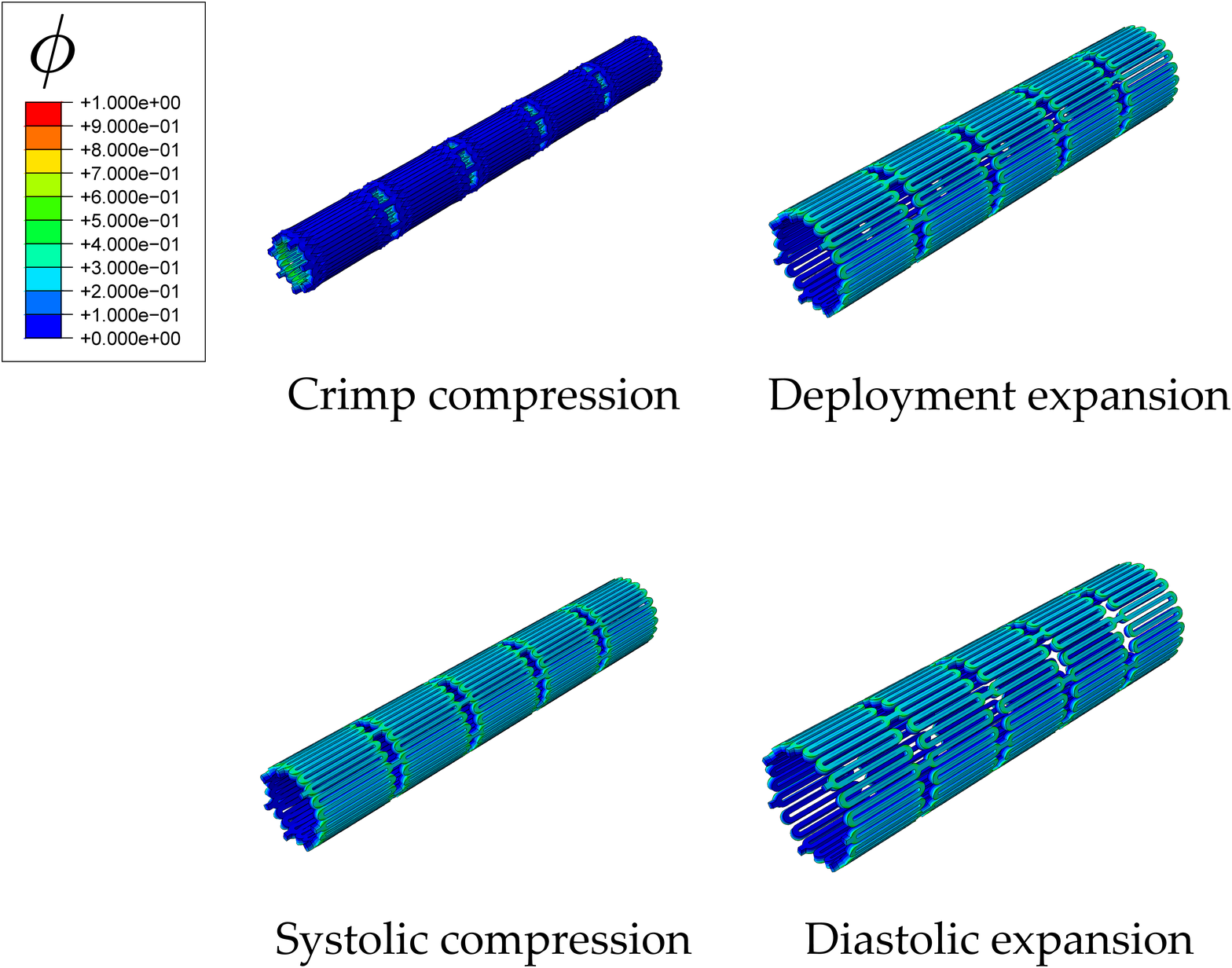}}
\caption{Phase field contours and deformed shape ($\times 8)$ of the SMA stent during crimp, deployment and the first systolic-diastolic pressure cycle.}
\label{fig:4cyclesStent}
\end{figure}

\begin{figure}[H]
\centering
\noindent\makebox[\textwidth]{%
\includegraphics[scale=0.07]{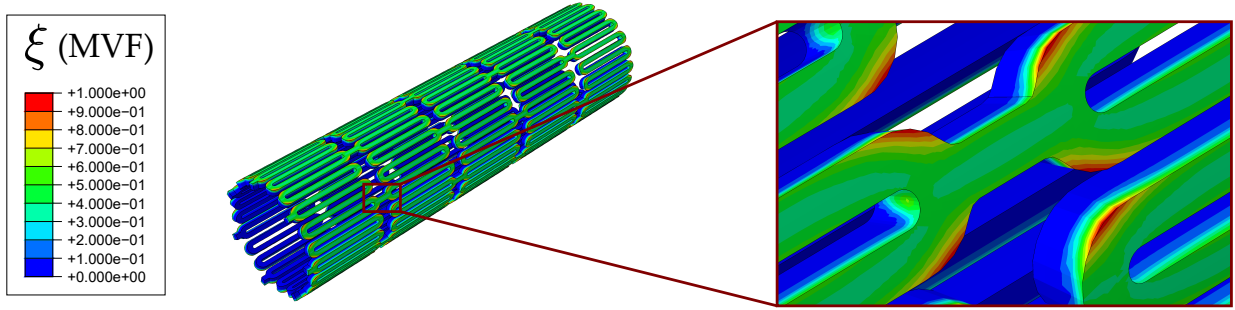}}
\caption{Contours of the martensitic volume fraction $\xi$ in the SMA stent during the deployment phase.}
\label{fig:PhaseTransStent}
\end{figure}

\begin{figure}[H]
\centering
\noindent\makebox[\textwidth]{%
\includegraphics[scale=0.07]{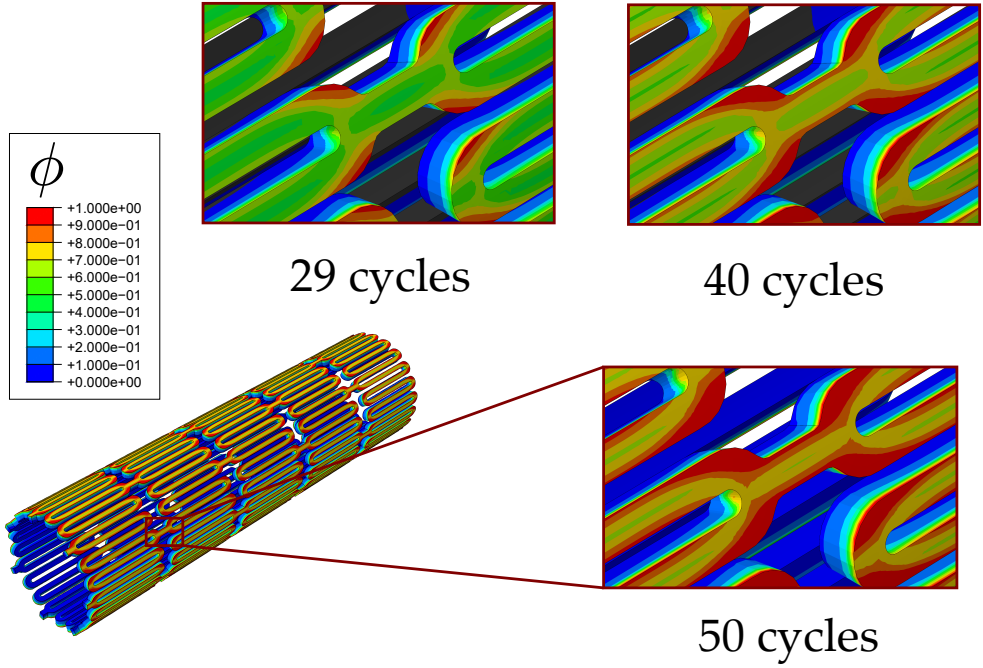}}
\caption{Contours of the phase field parameter $\phi$ in the SMA stent during (compression-expansion) systolic-diastolic cycling.}
\label{fig:FractureStent}
\end{figure}

The finite element results obtained are shown in Figs. \ref{fig:4cyclesStent}, \ref{fig:PhaseTransStent} and \ref{fig:FractureStent}, along with a video that is provided in the online version of this manuscript. Fig. \ref{fig:4cyclesStent} shows the deformed shape of the model during the four stages of the analysis: the crimp and expansion stages of the deployment phase, and the systolic compression and diastolic expansion stages associated with each pressure cycle. Contours of the phase field are shown with the deformed shape, revealing that values of $\phi$ of up to 0.56 are attained during the deployment phase. Thus, a significant amount of damage occurs during deployment but no cracking is observed. This is also the case for the phase transformation; most of it takes place during the deployment phase. The contours of martensitic volume fraction $\xi$ at the end of the crimp-expansion deployment process are shown in Fig. \ref{fig:PhaseTransStent}. 

The $\xi$ contours show that the martensite phase is localised in the edges of the stent struts. This is also the location where cracking initiates. While $\phi$ reaches high values during the deployment stage, 29 cycles of (compression-expansion) systolic-diastolic pressure are needed for cracks to initiate. As shown in Fig. \ref{fig:FractureStent}, these surface cracks initiate in the regions where phase transformation took place during stent deployment. The cracked region ($\phi=1$) extends with increasing the number of pressure cycles, and after 50 cycles it has extended over a significant part of the stent, including the bridging areas between struts. The evolution of the phase field over 6 cycles is shown in Video 1, provided in the online version of this manuscript. In summary, the example demonstrates that the framework presented here can be used to estimate the lifetime of medical stents for arbitrary geometries, boundary conditions and material properties.\\

\section{Conclusions}
\label{Sec:ConcludingRemarks}

We have presented the first phase field fracture formulation for Shape Memory Alloys (SMAs). The model can capture both fracture and fatigue damage and includes the main features inherent to the constitutive behaviour of SMAs; including superelasticity, shape memory effect, gradual phase transformations and the stress-sensitivity of the inelastic recoverable strain and of the phase diagram slope. From a fracture perspective, key features include the definition of a martensite volume fraction ($\xi$)-dependent fracture energy and the consideration of the AT1 and AT2 models in the constitutive definition of the crack density function. The theoretical model is numerically implemented using the finite element method, including both staggered and (quasi-Newton) monolithic schemes. The potential and robustness of the computational framework presented are demonstrated by addressing several paradigmatic 2D and 3D boundary value problems involving subcritical crack growth, unstable cracking, crack coalescence and fatigue crack growth of a nitinol stent. The main findings are:\\

(i) A stress-induced transformation zone develops in the vicinity of the crack tip and three distinct regions are observed: an inner martensite region with $r^{-1/2}$ singularity, an intermediate phase transformation region, and an outer austenite region with $r^{-1/2}$ singularity.\\

(ii) The stress-induced transformation phenomenon leads to inelastic energy dissipation and material toughening. This toughening effect is more significant at higher material strengths and lower temperatures. The use of a uniform toughness and the AT2 choice for the crack density function also increases crack growth resistance relative to a $\xi$-dependent fracture energy and the AT1 model, respectively.\\

(iii) Boundary value problems that favour the appearance of inelastic shear bands and large phase transformation regions can lead to notable subcritical crack propagation. Contrarily, where the phase transformation is confined to a small region fracture occurs in an unstable manner.\\

(iv) The coupling with very recent developments in phase field fatigue enables predicting the service life of SMA components in practical applications, as demonstrated with the analysis of a stent.\\

Potential extensions to the current framework include the consideration of plastic straining and of thermal loads.

\section{Acknowledgements}
\label{Sec:Acknowledgeoffunding}

The authors would like to acknowledge the assistance of Dr Darren J. Hartl (Texas A\&M University) in relation to the implementation of the SMA constitutive model. M. Simoes acknowledges insightful discussions with Dr Advenit Makaya (European Space Agency, ESA) and Dr Christopher Braithwaite (University of Cambridge), as well as financial support from the EPSRC (grant EP/R512461/1) and from the ESA (Contract no. 4000125861). E. Mart\'{\i}nez-Pa\~neda acknowledges financial support from the EPSRC (grants EP/R010161/1 and EP/R017727/1) and from the Royal Commission for the 1851 Exhibition (RF496/2018).






\bibliographystyle{elsarticle-num}
\bibliography{library}

\end{document}